\documentclass[twocolumn,twoside]{IEEEtran}

\usepackage[utf8]{inputenc} 
\usepackage[T1]{fontenc} 
\usepackage[cmex10]{amsmath} 

\usepackage{amssymb} 
\usepackage{mathtools} 
\usepackage{amsfonts} 
\usepackage{accents} 

\usepackage{mleftright}\mleftright 

\usepackage[dvips]{graphicx} 
\graphicspath{{./figures/}}
\DeclareGraphicsExtensions{.eps}

\usepackage[dvipsnames]{xcolor} 

\usepackage{bbm} 
\usepackage{xfrac} 
\usepackage{cancel} 
\usepackage{wrapfig} 
\usepackage{tensor} 

\usepackage{theorem}
\usepackage{cite}

\usepackage{url} 

\usepackage{psfrag} 

\usepackage[inline]{enumitem} 

\usepackage{array} 

\usepackage{tabu} 
\newcolumntype{M}[1]{>{\centering\arraybackslash}m{#1}}
\newcolumntype{R}[1]{>{\arraybackslash}m{#1}}
\newcolumntype{N}{@{}m{0pt}@{}}

\usepackage{aliascnt} 

\usepackage[hidelinks,linktocpage]{hyperref}

\usepackage{cleveref} 
\usepackage{xparse} 

\usepackage{ifthen} 

\crefname{section}{Section}{Sections}
\crefname{subsection}{Section}{Sections}
\crefname{equation}{Eq.}{Equations}
\crefname{enumi}{part}{parts}

\newtheorem{theorem}{Theorem}
\crefname{theorem}{Theorem}{Theorems}

\newaliascnt{lemma}{theorem}
\newtheorem{lemma}[lemma]{Lemma}
\aliascntresetthe{lemma}
\crefname{lemma}{Lemma}{Lemmas}

\newaliascnt{definition}{theorem}
\newtheorem{definition}[definition]{Definition}
\aliascntresetthe{definition}
\crefname{definition}{Definition}{Definitions}

\newaliascnt{corollary}{theorem}
\newtheorem{corollary}[corollary]{Corollary}
\aliascntresetthe{corollary}
\crefname{corollary}{Corollary}{Corollarys}

\newaliascnt{example}{theorem}
\newtheorem{example}[example]{Example}
\aliascntresetthe{example}
\crefname{example}{Example}{Examples}

\newaliascnt{claim}{theorem}

\aliascntresetthe{claim}
\crefname{claim}{Claim}{Claims}

\newaliascnt{conjecture}{theorem}

\aliascntresetthe{conjecture}
\crefname{conjecture}{Conjecture}{Conjectures}

\newaliascnt{question}{theorem}

\aliascntresetthe{question}
\crefname{question}{Question}{Questions}

\newaliascnt{oquestion}{theorem}

\aliascntresetthe{oquestion}
\crefname{oquestion}{Open Question}{Open Questions}

\theoremstyle{plain}
\theorembodyfont{\normalfont}

\newtheorem{cnstr}{Construction}

\crefname{cnstr}{Construction}{Constructions}

\crefname{step}{Step}{Steps}

\crefname{regime}{Regime}{Regimes}

\newtheorem{myalgo}{Algorithm}

\newenvironment{myalgorithm}{\begin{myalgo}}{\vspace*{-1\baselineskip}%
\hfill$\Box$\end{myalgo}}
\crefname{myalgo}{Algorithm}{Algorithms}

\newcounter{enumrom}
\renewcommand{\theenumrom}{(\roman{enumrom})}

\makeatletter
\renewcommand{\@endtheorem}{\endtrivlist}
\makeatother

\makeatletter
\renewcommand{\thefigure}{{\@arabic\c@figure}}
\renewcommand{\fnum@figure}{{\bf Figure\,\thefigure}}
\makeatother

\renewcommand{\leq}{\leqslant}
\renewcommand{\geq}{\geqslant}

\newcommand{\cC}{\mathcal{C}}
\newcommand{\cD}{\mathcal{D}}

\newcommand{\cT}{\mathcal{T}}

\renewcommand{\Bbb}{\mathbb}

\newcommand{\N}{{\Bbb N}}

\newcommand{\Z}{{\Bbb Z}}
\newcommand{\E}{{\Bbb E}}

\newcommand{\1}{\mathbbm{1}}

\DeclarePairedDelimiter\abs{\lvert}{\rvert}
\DeclarePairedDelimiter\ceilenv{\lceil}{\rceil}
\DeclarePairedDelimiter\floorenv{\lfloor}{\rfloor}
\DeclarePairedDelimiter\parenv{\lparen}{\rparen}

\DeclarePairedDelimiter\bracenv{\lbrace}{\rbrace}
\DeclarePairedDelimiterX\mathset[2]{\lbrace}{\rbrace}{#1 : #2}
\DeclarePairedDelimiterX\inner[2]{\langle}{\rangle}{#1 \mathrel{},\mathrel{} #2}
\DeclarePairedDelimiterX\condparenv[2]{(}{)}{#1 \mathrel{}\delimsize\vert\mathrel{} #2}

\DeclareDocumentCommand\norm{ o m }{
    \IfNoValueTF{#1}
        {\left\Vert#2\right\Vert}
        {\left\Vert#2\right\Vert_{#1}}
}

\DeclareDocumentCommand\der{ o m o }{
    \IfNoValueTF{#1}
        {
            \IfNoValueTF{#3}
                {\frac{d}{d{#2}}}
                {\frac{d{#3}}{d{#2}}}
        }
        {\parenv*{\frac{d}{d{#2}}}^{#1}\IfNoValueTF{#3}{}{#3}}
}
\DeclareDocumentCommand\partder{ o m m }{
    \IfNoValueTF{#1}
        {\frac{\partial{#3}}{\partial{#2}}}
        {\frac{\partial^{#1}{#3}}{{\partial{#2}}^{#1}}}
}
\DeclareDocumentCommand\df{ o m o }{
    d\IfNoValueTF{#1}{}{^{#1}}{#2}\IfNoValueTF{#3}{}{_{#3}}
}

\newcommand{\tends}[1]{\underset{#1}{\longrightarrow}}

\newcommand{\desc}[2]{\overset{#2}{\underset{#1}{\implies}}}

\newcommand{\deq}{\mathrel{\triangleq}}

\DeclareMathOperator{\GF}{GF}

\DeclareMathOperator{\irr}{Irr}

\DeclareMathOperator{\rt}{drt}

\DeclareMathOperator{\wt}{wt}

\DeclareMathOperator{\typ}{Typ}

\DeclareMathOperator{\poly}{poly}

\begin{document}
\title{Uncertainty of Reconstruction with List-Decoding from Uniform-Tandem-Duplication Noise}
\author{Yonatan Yehezkeally~\IEEEmembership{Member,~IEEE},
        \and Moshe Schwartz~\IEEEmembership{Senior~Member,~IEEE}
        \thanks{This work was presented in part at {ISIT}'2020.}%
        \thanks{This work was supported in part by the Israel Science Foundation (ISF) under grant no.~270/18.}%
        \thanks{Moshe Schwartz is with the School of Electrical and Computer Engineering, Ben-Gurion University of the Negev, Beer Sheva 8410501, Israel
  (e-mail: schwartz@ee.bgu.ac.il).}%
        \thanks{Yonatan Yehezkeally was with the School of Electrical and Computer Engineering, Ben-Gurion University of the Negev, Beer Sheva 8410501, Israel. He is now with the Institute for Communications Engineering, Technical University of Munich, 80333 Munich, Germany
  (e-mail: yonatan.yehezkeally@tum.de).}%
        \thanks{Copyright (c) 2021 IEEE. Personal use of this material is permitted. However, permission to use this material for any other purposes must be obtained from the IEEE by sending a request to pubs-permissions@ieee.org.}
}
\maketitle
\begin{abstract}
We propose a list-decoding scheme for reconstruction codes in the 
context of uniform-tandem-duplication noise, which can be viewed as 
an application of the associative memory model to this setting. We 
find the uncertainty associated with $m>2$ strings (where a previous 
paper considered $m=2$) in asymptotic terms, where code-words are 
taken from an error-correcting code. Thus, we find the trade-off 
between the design minimum distance, the number of errors, the 
acceptable list size and the resulting uncertainty, which corresponds 
to the required number of distinct retrieved outputs for successful 
reconstruction. It is therefore seen that by accepting list-decoding 
one may decrease coding redundancy, or the required number of reads, 
or both.
\end{abstract}


\section{Introduction}

\IEEEPARstart{W}{ith} recent improvements in DNA sequencing and
synthesis technologies, and the advent of CRISPR/Cas gene editing
technique \cite{Shi17}, the case for DNA as a data-storage medium,
specifically \textit{in-vivo}, is now stronger than ever before. It
offers a long-lasting and high-density alternative to current storage
media, particularly for archival purposes \cite{ChuGaoKos12}. 
Moreover, due to medical necessities, the technology required for 
data retrieval from DNA is highly unlikely to become obsolete, which 
as recent history shows, cannot be said of concurrent alternatives 
(e.g., the floppy disk, compact cassete, VHS tape, etc.).

In-vivo DNA storage has somewhat lower data density than in-vitro 
storage, but it provides a reliable and cost-effective propagation 
via replication, in addition to some protection to stored data 
(see \cite{Bal13,JaiFarSchBru17a,JaiFarSchBru17b} and references 
therein). 
It also has applications including watermarking genetically modified 
organisms \cite{AriOha04,HeiBar07,LisDauBruKliHamLeiWag12} or 
research material \cite{WonWonFoo03,JupFicSamQinFig10} and concealing 
sensitive information \cite{CleRisBan99}. However, mutations 
introduce a diverse set of potential errors, including symbol- or 
burst-substitution/insertions/deletion, and duplication (including 
tandem- and interspersed-duplication).

The effects of duplication errors, specifically, were studied in a 
number of recent works including \cite{JaiFarSchBru17a,JaiFarBru17,
SalGabSchDol17,MahVar17,LenJunWac18b,KovTan18a,LenWacYaa17b,
LenWacYaa19,Kov19b,TanYehFarSch20,TanFar19,TanFar20} among others. 
These works provided some implicit and explicit constructions for 
uniform-tandem-duplication codes, as well as some bounds. 
In \cite{YehSch19} the authors then argued that a classical 
error-correction coding approach is sub-optimal for the application, 
as it does not take advantage of the cost-effective data replication 
offered inherently by the medium of in-vivo DNA; instead, it was 
shown that re-framing the problem as a \emph{reconstruction} scheme 
\cite{Lev01} reduces the redundancy required for any fixed number of 
duplication errors. In this setting, several (distinct) noisy channel 
outputs are assumed to be available to the decoder. Since its 
introduction, several applications of the reconstruction problem to 
storage technologies were found \cite{CasBla11,YaaBruSie16,
CheKiaVarVuYaa18,YaaBru19}. Of these, \cite{YaaBru19} in particular 
extended the reconstruction model to \emph{associative memory}, where 
one retrieves the set of all entries (or code-words) 
\emph{associated} with every element of a given set. For a given size 
of entry set, the maximal number of entries being possibly associated 
with all of them was dubbed the \emph{uncertainty} of the memory.

Study of this extended model for in-vivo DNA data storage is 
motivated by a list-decoding reconstruction scheme, whereby tolerance 
for decoding a list of possible inputs, given multiple channel 
outputs, enables coding with a lower minimum distance, thereby 
reducing the redundancy of the code. Alternatively, given the same 
code, it allows reducing the number of required outputs for 
reconstruction.

This paper focuses on uniform tandem-duplication noise; i.e., we 
assume throughout that the length of duplication window is fixed. In 
practical applications, a more complex model where that length is 
permitted to belong to some set, or perhaps is simply bounded, is 
more realistic; however, we focus on this model as a step towards that 
end. Our main goal is to analyze the uncertainty associated with codes 
which are subsets of a typical set of strings (consisting of most 
strings in $\Sigma^n$, a definition which is made precise in 
\cref{lem:typ}) as a function of the acceptable list size~$m$ and 
code minimum distance~$d$. In our analysis, the number of tandem 
repeats~$t$ which channel outputs undergo is fixed.

The paper is organized as follows: In \cref{sec:contrib} we 
describe the main contribution of this paper, put it in context of 
related works, and discuss possible directions for future study. In 
\cref{sec:preliminaries} we present notations and definitions. 
Then, in \cref{sec:typ}, we find the uncertainty of the 
aforementioned typical set in asymptotic form, and develop an 
efficient decoding scheme. We then extend and repeat our analysis in 
\cref{sec:ecc} for error-correcting codes contained in that 
typical set.

\section{Related works and main contribution}\label{sec:contrib}

Associative memory was discussed in~\cite{YaaBru19}, where items are
retrieved by association with other items; the human mind seems to
operate in this fashion, one concept bringing up memories of other,
related, concepts or events. The more items one considers together,
the smaller the set of items associated with all of them. 
More precisely, one defines the uncertainty of an associative memory 
as the largest possible size of set $N(m)$ whose members are 
associated with all elements of an $m$-subset of the memory code-book.

This model is a generalization of the reconstruction problem posed by
Levenshtein in \cite{Lev01}, wherein a transmission model is assumed
with the decoder receiving multiple channel outputs of the same
input. $N$ is then the largest size of intersection of balls of radius
$t$ about two distinct code-words, where at most $t$ errors are
assumed to have occurred in each transmission; if $N+1$ outputs are
available to the decoder, the correct input can be deduced.

This can be viewed as a reduction of the associative memory model to
the case of $m=2$, allowing a precise reconstruction of the unique
($m-1=1$) input. When $m>2$, the decoder seeing $N(m)+1$ channel
outputs can only unambiguously infer which list of $l<m$ code-words
contains the correct input; thus, a list-decoding model is suggested.

In \cite{YehSch19} the authors studied the reconstruction problem for
uniform-tandem-duplication noise, which is applicable to in-vivo DNA
data storage. An uncertainty which is sub-linear in the message length
was assumed (as it represents the number of distinct reads required
for decoding), and it was shown that the redundancy required for
unique reconstruction was ${(t-1)\log_q(n)} + O(1)$ (compared to the 
${t\log_q(n)} + O(1)$ redundancy required for unique decoding from a 
single output \cite{KovTan18a,LenJunWac18b}), where $n$ is the message 
length, $t$ the number of errors, and $q$ the alphabet size.

In this paper, we apply the associative memory model from
\cite{YaaBru19} (where binary vectors with the Hamming distance were
considered) to the setting of uniform-tandem-duplication noise in
finite strings, i.e., we consider list-decoding instead of a unique
reconstruction. We shall restrict our attention to code-books
contained in a typical subspace, asymptotically achieving the full
space size.

Our goal is to find the trade-off between the code redundancy, the 
number of tandem-duplication errors, the uncertainty, and the decoded 
list size. We find the asymptotic behavior, as the message length~$n$ 
grows, of the uncertainty, or required number of reads (more 
precisely, that number minus one) $N$, where it is viewed as a 
function of the list size (plus one)~$m$, the design minimum 
distance~$d$, and the number of tandem-duplication errors~$t$. Our 
main contribution (see \cref{cor:tradeoff}) can informally be 
summarized in 
\[
\log_n N + \ceilenv*{\log_n(m)} + d = t + \epsilon + o(1),
\]
where $\epsilon\in \bracenv*{0,1}$ is a non-increasing function 
of~$m$, which we find. Thus, such a trade-off is established.

This can be seen as an extension to the results in \cite{YehSch19},
where unique reconstruction ($m=2$) was required, and it was seen that
coding with minimum distance $d=t$ enables sub-linear uncertainty
(i.e., $\log_n(N) = o(1)$).

In conclusion, we show that list-decoding is not only theoretically 
feasible, but may be efficiently performed. This is done using an 
isometric transform to integer vectors, and by utilizing combination 
generators; efficient list-decoding algorithms are developed, given 
a sufficient number of distinct channel outputs. If the code-book is 
restricted, then this task is reduced to that of decoding an 
error-correcting code.

In the future, we believe that a study of reconstruction schemes, 
with or without list-decoding, is of interest with other error models 
which affect in-vivo DNA data storage; related models to uniform 
tandem-duplication noise, which have recently been studied on their 
own and may now be easier to analyze in that setting, and therefore 
are a logical first step in this direction, may be bounded 
tandem-duplication (see, e.g., \cite{JaiFarSchBru17a,JaiFarBru17,
Kov19b}) or combined uniform-tandem-duplication and substitution noise 
\cite{TanYehFarSch20,TanFar19,TanFar20}.

\section{Preliminaries}\label{sec:preliminaries}

Let $\Sigma^*$ denote the set of finite strings over an alphabet 
$\Sigma$, which is assumed to be a finite unital ring of size $q$ 
(e.g., $\Z_q$, or when $q$ is a prime power, $\GF(q)$).

The length of a string $x\in\Sigma^*$ is denoted $\abs*{x}$, and the 
concatenation of $x,y\in\Sigma^*$ is denoted $xy$. A 
tandem-duplication (or tandem repeat) of fixed duplication-window 
length $k$ (thus, \emph{uniform} tandem-duplication noise) at index 
$i$ is defined as follows, for $a\in\Sigma^*$ such that $a=xyz$, 
$x,y,z\in\Sigma^*$, $\abs*{x}=i$ and 
$\abs*{y}=k$: 
\[
\cT_i(a) \deq xyyz.
\]
Thus, uniform tandem-duplication noise with duplication-window 
length~$k$ acts only on strings of length $\geq k$, which we denote 
$\Sigma^{\geq k}$. In order to simplify our analysis, we assume 
throughout the paper that $k\geq 2$.

If $y\in\Sigma^{\geq k}$ can be derived from $x\in\Sigma^{\geq k}$ by 
a sequence of tandem repeats, i.e., if there exist $i_1,\ldots,i_t$ 
such that 
\[
y = \cT_{i_t}\parenv*{\cdots \cT_{i_1}(x)},
\]
then $y$ is called a \emph{$t$-descendant} (or simply 
\emph{descendant}) of $x$ (vice versa, $x$ is an \emph{ancestor} of 
$y$), and we denote ${x\desc{\mathclap{}}{t}y}$. We say that $x$ is a 
$0$-descendant of itself. If $t=1$ we denote $x\Longrightarrow y$. 
Where the number of repeats is unknown or irrelevant, we may denote 
$x\overset{\mathclap{*}}{\Longrightarrow} y$. We define the set of 
$t$-descendants of $x$ as 
\[
D^t(x) \deq \mathset*{y\in\Sigma^*}{x\desc{}{t}y},
\]
and the \emph{descendant cone} of $x$ as 
\[
D^*(x) \deq \mathset*{y\in\Sigma^*}{x\desc{}{*}y} 
= \bigcup_{t=0}^\infty D^t(x).
\]

If there exists no $z\neq x$ such that $z\desc{}{*}x$, we say that $x$
is \emph{irreducible}. The set of irreducible strings of length $n$ is
denoted $\irr(n)$. It can be shown (see, e.g., \cite{JaiFarSchBru17a})
that for all $y\in\Sigma^{\geq k}$ there exists a unique irreducible
$x$, called the \emph{duplication root} of $y$ and denoted $\rt(y)$,
such that $y\in D^*(x)$. This induces a partition of $\Sigma^{\geq k}$
into descendant cones; i.e., it induces an equivalence relation, 
denoted herein $\sim_k$.

A useful tool in studying uniform tandem-duplication noise is the
\emph{discrete derivative} $\phi$ defined for $x\in\Sigma^{\geq k}$:
\[
\phi(x) \deq \hat{\phi}(x) \bar{\phi}(x),
\]
where
\begin{align*}
\hat{\phi}(x) &\deq x(1),x(2),\ldots,x(k), \\
\bar{\phi}(x) &\deq x(k+1)-x(1),\ldots,x(\abs*{x}) - x(\abs*{x}-k).
\end{align*}
Here, $x(i)$ denote the $i^{\text{th}}$ letter of the string $x$; 
Note that $\hat{\phi}(x),\bar{\phi}(x)$, and consequently $\phi(x)$, 
are themselves strings in $\Sigma^*$. 
As seen, e.g., in \cite{JaiFarSchBru17a}, $\phi$ is injective, and if
$\bar{\phi}(x) = uv$ for $u,v\in\Sigma^*$, $\abs*{u}=i$, then
$\bar{\phi}\parenv*{\cT_i(x)} = u 0^k v$. This was used in
\cite{YehSch19} to define the function $\psi_x\colon
D^*(x)\to\N^{w+1}$ by
\[
\psi_x(y) \deq (\floorenv*{u(1)/k},\ldots,\floorenv*{u(w+1)/k}),
\]
if
\[
\bar{\phi}(y) = 0^{u(1)} a_1 0^{u(2)} \ldots a_w 0^{u(w+1)},
\]
where $w = \wt(\bar{\phi}(x))$ and
$a_1\ldots,a_w\in\Sigma\setminus\bracenv*{0}$. It was shown that
$\psi_x$ is a poset isomorphy, where $D^*(x)$ is ordered with
$\desc{}{*}$ and $\N^{w+1}$ with the product order, which we denote 
by $\leq$. 
Further, when considering $\N^{w+1}$ as a poset with the product 
order, we shall use the notations $\vee,\wedge$ for the 
\emph{supremum} and \emph{infimum}, respectively; these are also 
the coordinatewise \emph{maximum} and \emph{minimum}, respectively.

A metric can be defined on $D^r(x)$ for each $r$ (in particular, but
not necessarily, when $x$ is irreducible) in the following way:
\begin{definition}\label{def:dist}
For any $r\in\N$, $x\in\Sigma^{\geq k}$, and $y_1,y_2\in D^r(x)$, we 
define
\[
d(y_1,y_2) 
\deq \min\mathset*{t\in\N}{D^t(y_1)\cap D^t(y_2)\neq\emptyset}.
\]
\end{definition}
It is seen in \cite{JaiFarSchBru17a} that this is well defined, in
the sense that there does exist such $t$, for $y_1,y_2\in D^r(x)$,
such that $D^t(y_1)\cap D^t(y_2) \neq\emptyset$.

If we define on $\N^{w+1}$ the $1$-norm
\[
\norm[1]{u}\deq \sum_{i=1}^{w+1}u(i),
\]
and metric
\[
d_1(u,v) \deq \tfrac{1}{2}\norm[1]{u-v},
\]
then $\psi_x$ is also an isometry (see \cite{YehSch19}) between 
$D^r(x)$, for each $r$, and its image in $\N^{w+1}$, which is the 
simplex 
\[
\Delta^w_r \deq \mathset*{u\in\N^{w+1}}{\norm[1]{u}=r} 
= \psi_x\parenv*{D^r(x)}.
\]
Here, $\psi_x\parenv*{D^r(x)}$ is the image of $\psi_x$; in more 
generality, for any code $C\subseteq D^r(x)$ we let $\psi_x(C) \deq 
\mathset*{\psi_x(y)}{y\in C}$.

To simplify analysis in the $\N^{w+1}$ domain, we make the following 
notation:
\begin{definition}
For $w,r,s\in\N$ and $u\in\Delta^w_{r+s}$, denote the 
\emph{lower-bounds set} 
\[
A_r(u) \deq \mathset*{v\in\Delta^w_r}{v\leq u}.
\]
\end{definition}

The focus of this paper is to find the uncertainty, after $t$~tandem 
repeats, as a function of the acceptable list size~$m$. This is made 
precise by the following definition.

\begin{definition}\label{def:unc}
Given $n,t\in\N$ and $x_1,\ldots,x_m\in\Sigma^n$, we define 
\[
S_t\parenv*{x_1,\ldots,x_m} \deq \bigcap_{i=1}^m D^t(x_i).
\]
Then, the \emph{uncertainty} associated with a code $C\subseteq 
\Sigma^n$ is 
\[
N_t(m,C) \deq \max_{\substack{x_1,\ldots,x_m\in C \\ x_i\neq x_j}} 
\abs*{S_t\parenv*{x_1,\ldots,x_m}}.
\]

Correspondingly, for $w,r\in\N$ and $u_1,\ldots,u_m\in\Delta^w_r$ we 
define 
\begin{align*}
\bar{S}_t\parenv*{u_1,\ldots,u_m} &\deq \bigcap_{i=1}^m 
\mathset*{v\in\N^{w+1}}{v\geq u_i,\,\norm[1]{v-u_i}=t} ; \\
\bar{N}_t(m,w,r) &\deq \max_{u_1,\ldots,u_m\in\Delta^w_r} 
\abs*{\bar{S}_t\parenv*{u_1,\ldots,u_m}}.
\end{align*}
\end{definition}

In the next section we describe a typical set of strings in 
$\Sigma^n$, then by ascertaining $\bar{N}_t(m,w,r)$ for that set we 
find an asymptotic expression (in the string length $n$) for the 
uncertainty associated with that set, as a function of $m$.

Finally, in our analysis we shall use the following asymptotic 
notation: for two sequences $a_n, b_n$ we say that $a_n\sim b_n$ if 
$a_n=b_n(1+o(1))$.

\section{Typical set}\label{sec:typ}

We observe that the sets introduced in the previous section have many
parameters. 
A complete combinatorial analysis of those would be encumbered by 
extreme cases which occur in a vanishingly small fraction of the 
space; analysis of these cases is therefore not only more 
challenging, but also less enlightening. Since our main goal is an 
asymptotic analysis, we proceed by eliminating those rare 
pathological cases, and focus on the common typical ones. In 
particular, we would like to limit our attention to strings 
$x\in\Sigma^n$ for which the Hamming weight of $\bar{\phi}(x)$ and the 
$1$-norm of $\psi_{\rt(x)}(x)$, as well as the difference between 
them, are asymptotically linearly proportional to the string 
length~$n$. 
Those strings would form the code which we study. Thus, we start by 
presenting in the following lemma the code~$C$ for which it shall be 
our goal to find $N_t(m,C)$.

\begin{lemma}\label{lem:typ}
Define the family of codes 
\[
\typ^n \deq \mathset*{x\in\Sigma^n}{
\begin{smallmatrix}
\abs*{w(x)-\frac{q-1}{q}(n-k)} < n^{3/4} \\
\abs*{r(x)-\frac{q-1}{q(q^k-1)}(n-k)} < 2 n^{3/4}
\end{smallmatrix}
},
\]
where $w(x)\deq \wt_H\parenv*{\bar{\phi}(x)}$ and 
$r(x) \deq \norm[1]{\psi_{\rt(x)}(x)}$. Then for sufficiently 
large~$n$: 
\[
\frac{\abs*{\typ^n}}{\abs*{\Sigma^n}} 
\geq 1-4e^{-\sqrt{n}/2} \tends{n\to\infty} 1.
\]
\end{lemma}
\begin{IEEEproof}
We note that if $x,y\in\Sigma^n$ differ only in a single coordinate,
then $\abs*{w(x)-w(y)},\abs*{r(x)-r(y)}\leq 2$. If the $x(i)$'s are
thought of as independent and uniformly distributed random variables
on $\Sigma$, then by McDiarmid's inequality \cite{Doo40} we have 
\begin{align*}
&\tfrac{1}{\abs*{\Sigma^n}} 
\abs*{\mathset*{x\in\Sigma^n}{\abs*{w(x)-\E[w(x)]}\geq n^{3/4}}} 
\leq 2 e^{-\sqrt{n}/2}, \\
&\tfrac{1}{\abs*{\Sigma^n}} 
\abs*{\mathset*{x\in\Sigma^n}{\abs*{r(x)-\E[r(x)]} \geq n^{3/4}}} 
\leq 2 e^{-\sqrt{n}/2}.
\end{align*}
Further note that if $\E[r(x)] = \alpha(n-k) + o(n^{3/4})$ then for 
large enough $n$ we also have 
\[
\tfrac{1}{\abs*{\Sigma^n}} 
\abs*{\mathset*{x\in\Sigma^n}{\abs*{r(x)-\alpha (n-k)} 
\geq 2n^{3/4}}} \leq 2 e^{-\sqrt{n}/2},
\]
and hence
\[
\tfrac{1}{\abs*{\Sigma^n}}\abs*{\mathset*{x\in\Sigma^n}{
\begin{smallmatrix}
\abs*{w(x)-\E[w(x)]} < n^{3/4} \\
\abs*{r(x)-\alpha (n-k)} < 2n^{3/4}
\end{smallmatrix}}
} \geq 1 - 4 e^{-\sqrt{n}/2}.
\]

Next, note that $u(i)\deq\parenv*{\bar{\phi}(x)}(i)$ are also 
independent and uniformly distributed. Define the indicator functions 
$a(i)\deq\1_{\bracenv*{u(i)\neq 0}}$. Clearly 
\[
\E[w(x)] = \sum_{i=1}^{n-k} \E[a(i)] = \sum_{i=1}^{n-k}
\Pr(u(i)\neq 0) = \tfrac{q-1}{q}(n-k).
\]

See the Appendix 
for proof that $\E[r(x)] = \frac{q-1}{q(q^k-1)}(n-k) + O(1)$, which 
concludes the proof.
\end{IEEEproof}

We remark that a similar concentration result (for $w(x)$ and
$\wt_H(\psi_{\rt(x)}(x))$ instead of $r(x)$) was derived in
\cite[Lem.~3]{KovTan18a} using a different approach.

Before analyzing the uncertainty $N_t(m,\typ^n)$, we note that the 
process of list-decoding given sufficiently many ($N_t(m,\typ^n)+1$) 
distinct strings in $\Sigma^{n+kt}$, i.e., finding $x_1,\ldots,x_l 
\in\typ^n$, $l<m$, such that these strings lie in 
$
S_t(x_1,\ldots,x_l) \setminus 
\bigcup_{x\in\typ^n\setminus\bracenv*{x_1,\ldots,x_l}}D^t(x)
$, 
is straightforward:
\begin{myalgorithm}\label{alg:decode}
Denote $N\deq N_t(m,\typ^n)$ and assume as input distinct
$y_1,\ldots,y_{N+1} \in \Sigma^{n+kt}$ such that there exists
$x\in\typ^n$ satisfying $y_1,\ldots,y_{N+1} \in D^t(x)$.
(Note that, when such $x$ exists, $\rt(x)$ may be determined 
by, e.g., $\rt(x)=\rt(y_1)$; hence, $\psi_{\rt(x)}$ in particular may 
be used at will.)
\begin{enumerate}
\item\label{alg:decode:1} 
Apply $\psi_{\rt(y_1)}$ to map them to $v_1,\ldots,v_{N+1} \in 
\Delta^w_{r+t}$, where $w=\wt\parenv*{\bar{\phi}(\rt(y_1))}$ and
$r=\norm[1]{\psi_{\rt(y_1)}(y_1)}-t$; note that prior computation of 
$\rt(y_1)$ is not required to perform this mapping, and that it may 
be found as a byproduct of finding any $v_i$.

\item\label{alg:decode:2}
Find $u\deq \bigwedge_{i=1}^{N+1} v_i \in
\Delta^w_{r'}$ by calculating the minimum over each coordinate.

\item\label{alg:decode:3}
Calculate $A_r(u)$.

\item\label{alg:decode:4}
Return $\psi_{\rt(y_1)}^{-1}\parenv*{A_r(u)}$ as a list.
\end{enumerate}
\end{myalgorithm}

We defer proving the validity of \cref{alg:decode} to the end of 
the section, since an asymptotic evaluation of its run-time 
complexity involves analysis of $N_t(m,\typ^n)$, which we shall next 
tend to; before doing so, however, we shall present an example of the 
application of \cref{alg:decode}.

\begin{example}\label{xmpl:alg:decode}
Let $q=3, k=2$, and take $n=11, t=3$. We shall read multiple distinct 
elements of $D^t(x)$ for some unknown $x\in\typ^n$, and would like to 
decode a list of strings in $\typ^n$ of which $x$ is a member.

The first read we make is 
\[
y_1 = 10101012122222222.
\]
(This suffices to determine $\rt(y_1)=\rt(x)=10122$.)

Further, Suppose that we are willing to accept a list of size at most 
$3$, and therefore set $m=4$. Observing that 
\[
\bar{\phi}(y_1) = 000002001000000, 
\]
and consequently 
\[
w=\wt_H(\bar{\phi}(y_1))=2;
\quad
r = \norm[1]{\psi_{\rt(y_1)}(\bar{\phi}(y_1))} - t = 3,
\] 
it happens to be the case that $4$ distinct reads will suffice for 
this purpose.

(The reader referring back to this example, after having read the 
analysis following it, will note that \cref{lem:spread} and 
\cref{xmpl:mu_2} establish that $\mu(w,r,1)=3$ and $\mu(w,r,2)=4$, 
respectively; consequently, \cref{cor:mu} then implies 
$\sigma(4,w,r)=2$, and \cref{cor:N-by-sup} implies that 
$\bar{N}_t(m,w,r)=3$.

It is also of interest to note that, indeed, $x\in\typ^n 
= \typ^{11}$.)

We therefore make $3$ additional distinct reads of $D^t(x)$, 
obtaining:
\begin{align*}
y_2 &= 10101010122222222,\\
y_3 &= 10101012222222222,\\
y_4 &= 10101012121222222.
\end{align*}
We can now find 
\begin{align*}
\bar{\phi}(y_1) &= 000002001000000 = 0^{1+2k} 2 0^{k} 1 0^{3k},\\
\bar{\phi}(y_2) &= 000000021000000 = 0^{1+3k} 2 0^{0} 1 0^{3k},\\
\bar{\phi}(y_3) &= 000002100000000 = 0^{1+2k} 2 0^{0} 1 0^{4k},\\
\bar{\phi}(y_4) &= 000002000010000 = 0^{1+2k} 2 0^{2k} 1 0^{2k},\\
\end{align*}
which may be more succinctly represented by 
\begin{align*}
v_1 &= \psi_{\rt(y_1)}(y_1) = (2,1,3),\\
v_2 &= \psi_{\rt(y_1)}(y_2) = (3,0,3),\\
v_3 &= \psi_{\rt(y_1)}(y_3) = (2,0,4),\\
v_4 &= \psi_{\rt(y_1)}(y_4) = (2,2,2).
\end{align*}
This concludes \hyperref[alg:decode:1]{Step \ref*{alg:decode:1}}. 
The coordinatewise minimum required in 
\hyperref[alg:decode:2]{Step \ref*{alg:decode:2}} is therefore 
\[
u = (2,0,2),
\]
and for \hyperref[alg:decode:3]{Step \ref*{alg:decode:3}} we find 
$A_r(u) = A_3((2,0,2)) = \bracenv*{u_1,u_2}$, where 
\begin{align*}
u_1 &= (1,0,2),\\
u_2 &= (2,0,1).
\end{align*}

For \hyperref[alg:decode:4]{Step \ref*{alg:decode:4}}, we therefore 
find $x_i = \psi_{\rt(y_1)}^{-1}(u_i)$, $i\in\bracenv*{1,2}$, by
\begin{align*}
\bar{\phi}(x_1) &= 000210000 = 0^{1+k} 2 0^{k} 1 0^{2k},\\
\bar{\phi}(x_2) &= 000002100 = 0^{1+2k} 2 0^{k} 1 0^{k},
\end{align*}
and therefore 
\begin{align*}
x_1 &= 10101012222,\\
x_2 &= 10101010122.
\end{align*}
We note that the algorithm produced a list of size $2$, smaller than 
the design requirements; its guarantee of list size relies on maximal 
lower-bounds-set size, hence specific examples may well produce 
shorter lists.
\end{example}

Next, for $\typ^n$ we show that the uncertainty can be calculated by 
$\bar{N}_t$, which provides an expression we may more easily analyze.
\begin{lemma}\label{lem:intrsct-smplx}
For $C\subseteq\Sigma^n$, there exist $x\in\irr$ and 
$u_1,\ldots,u_m\in \psi_x(C\cap D^*(x))$ such that 
\[
N_t(m,C) = \abs*{\bar{S}_t\parenv*{u_1,\ldots,u_m}}.
\]
\end{lemma}
\begin{IEEEproof}
Take $x_1,\ldots,x_m\in C$ such that 
$\abs*{S_t\parenv*{x_1,\ldots,x_m}} \linebreak= N_t(m,C)$, and note that if 
there exist $x_i\not\sim_k x_j$, then $S_t\parenv*{x_1,\ldots,x_m} 
= \emptyset$, in contradiction. Hence there exists $x = 
\rt\parenv*{\bracenv*{x_1,\ldots,x_m}}$. The claim now follows from 
the isometry $\psi_x$, i.e., $\psi_x\parenv*{S_t(x_1,\ldots,x_m)} 
= \bar{S}_t\parenv*{\psi_x(x_1),\ldots,\psi_x(x_m)}$.
\end{IEEEproof}

\begin{corollary}\label{cor:bar-nobar}
For $k\geq 2$ and sufficiently large~$n$,
\begin{align*}
& N_t(m,\typ^n) =\\
&\qquad = \max\mathset*{\bar{N}_t(m,w,r)}{
\begin{smallmatrix}
\abs*{w-\frac{q-1}{q}(n-k)} < n^{3/4} \\
\abs*{r-\frac{q-1}{q(q^k-1)}(n-k)} < 2 n^{3/4}
\end{smallmatrix}
}.
\end{align*}
\end{corollary}
\begin{IEEEproof}
For convenience, we denote 
\[
M \deq \max\mathset*{\bar{N}_t(m,w,r)}{
\begin{smallmatrix}
\abs*{w-\frac{q-1}{q}(n-k)} < n^{3/4} \\
\abs*{r-\frac{q-1}{q(q^k-1)}(n-k)} < 2 n^{3/4}
\end{smallmatrix}
}.
\]
By \cref{lem:intrsct-smplx} we have $x\in\irr$ and $u_1,\ldots,u_m 
\in \psi_x(\typ^n\cap D^*(x))$ such that 
$
N_t(m,C) = \abs*{\bar{S}_t\parenv*{u_1,\ldots,u_m}}
$.
If we take $y_1,\ldots,y_m\in\typ^n\cap D^*(x)$ such that $\psi_x(y_i) 
= u_i$ for all $i\in\bracenv*{1,\ldots,m}$, then we have $w(y_1) = 
\ldots = w(y_m) = w(x)$. Furthermore, it follows from $\abs*{y_1} 
= \ldots = \abs*{y_m} = n$ that 
$r(y_1)=\ldots=r(y_m)$. Denote 
therefore $w\deq w(y_1)$ and $r\deq r(y_1)$, and since $y_1,\ldots, 
y_m \in\typ^n$ we have 
\[
\abs*{w-\tfrac{q-1}{q}(n-k)} < n^{3/4};
\quad 
\abs*{r-\tfrac{q-1}{q(q^k-1)}(n-k)} < 2 n^{3/4}.
\] 
Therefore, noting that $\bar{N}_t(m,w,r) \geq 
\abs*{\bar{S}_t\parenv*{u_1,\ldots,u_m}} = N_t(m,C)$, we conclude 
that $N_t(m,C)\leq M$.

To show the the other direction, note that for every pair $w,r$ 
satisfying 
\[
\abs*{w-\tfrac{q-1}{q}(n-k)} < n^{3/4};
\quad 
\abs*{r-\tfrac{q-1}{q(q^k-1)}(n-k)} < 2 n^{3/4},
\]
there exists $x\in\irr(n-kr)$ (so that $D^r(x)\subseteq\typ^n$) for 
which $w(x)=w$. This follows from counting the required number of 
zeros in $\bar{\phi}(x)$ for such~$x$, which is $n-(1+r)k-w$; for 
large enough~$n$ this number is positive and no greater than $(k-1)
(w+1)$. Hence, after arbitrarily choosing $w$ non-zero elements of 
$\Sigma$, we may pad them with runs of $k-1$ zeros or less (obtaining 
a string in $\bar{\phi}(\irr)$) to achieve a total length of 
$n-(1+r)k$. The derived string is $\bar{\phi}(x)$ for the desired 
$x\in\irr(n-rk)$, and by again arbitrarily choosing any $k$ elements 
of $\Sigma$ for the role of $\hat{\phi}(x)$, we may indeed find $x$ as 
desired.

Now, taking $w,r$ in the required ranges such that 
$M = \bar{N}_t(m,w,r)$, $u_1,\ldots,u_m\in\Delta^w_r$ such that 
$\abs*{\bar{S}_t\parenv*{u_1,\ldots,u_m}} = \bar{N}_t(m,w,r) = M$,
and $x\in\irr(n-rk)$ as described, we may find $y_1,\ldots,y_m\in 
\typ^n\cap D^*(x)$ by defining $y_i\deq \psi_x^{-1}(u_i)$ for all 
$i\in\bracenv*{1,\ldots,m}$, and note 
$N_t(m,C) \geq \abs*{S_t\parenv*{y_1,\ldots,y_m}} 
= \abs*{\bar{S}_t\parenv*{u_1,\ldots,u_m}} = M$.
\end{IEEEproof}

Hence, the quantity one needs to assess is $\bar{N}_t(m,w,r)$. We do
that next by exploiting the lattice structure of $\N^{w+1}$, and
introducing the connection to supremum height and lower-bound-set size
in that lattice.

\begin{lemma}\label{lem:intersection-sup}
Given $u_1,\ldots,u_m\in\Delta^w_r$, denote $u\deq \bigvee_{i=1}^m
u_i$. Then,
\[
\abs*{\bar{S}_t\parenv*{u_1,\ldots,u_m}} =
\begin{cases}
  0 & \norm[1]{u}>r+t,\\
  \binom{w+t+r-\norm[1]{u}}{w} & \text{otherwise.}
\end{cases}
\]
\end{lemma}
\begin{IEEEproof}
The proposition follows from the lattice structure of $\N^{w+1}$, 
i.e.,
\[
\bar{S}_t\parenv*{u_1,\ldots,u_m} 
=\mathset*{v\in\N^{w+1}}{v\geq u,\,\norm[1]{v-u_1}=t}.
\]
When $\norm[1]{u}-r = \norm[1]{u-u_1} > t$, then, the set is empty. 
Otherwise, the size of the set corresponds to the number of ways to 
distribute $t-\norm[1]{u-u_1} = t-(\norm[1]{u}-r)$ balls into $w+1$ 
bins.
\end{IEEEproof}

\begin{definition}\label{dfn:minsup-max}
Denote for $m,w,r\in\N$ the \emph{minimum supremum height} 
\[
\sigma(m,w,r) \deq 
\min_{u_1,\ldots,u_m\in\Delta^w_r} 
\norm[1]{\left.\bigvee\right._{i=1}^m u_i} - r.
\]
Conversely, for $w,r,s\in\N$ denote the \emph{maximal 
lower-bounds-set size} 
\[
\mu(w,r,s) \deq \max\mathset*{\abs*{A_r(u)}}{u\in\Delta^w_{r+s}}.
\]
(Recall that $A_r(u)$ is the lower-bounds set of $u$.)
\end{definition}

\begin{corollary}\label{cor:N-by-sup}
$\bar{N}_t(m,w,r) = \binom{w+t-\sigma(m,w,r)}{w}$.
\end{corollary}
\begin{IEEEproof}
The proposition follows from \cref{lem:intersection-sup}.
\end{IEEEproof}

It is therefore seen that the main task is to find or estimate the
minimum supremum height. We next show the duality between
$\sigma(m,w,r)$ and $\mu(w,r,s)$, which we shall use to calculate the
former.

\begin{lemma}\label{lem:sigma-mu}
Take $w,r,s\in\N$. If $s\geq wr$ then 
\[
\mu(w,r,s) = \abs*{\Delta^w_r} = \binom{r+w}{r} \quad\text{and}\quad 
\sigma(\abs*{\Delta^w_r},w,r) = wr.
\]
For $s < wr$ we have
\[
\sigma(\mu(w,r,s),w,r) = s.
\]
\end{lemma}
\begin{IEEEproof}
The first part of the proposition is justified by
$(r,r,\ldots,r)\in\Delta^w_{(w+1)r}$.

For the second, take $u\in\Delta^w_{r+s}$ satisfying
$\abs*{A_r(u)}=\mu(w,r,s)$. Since $\bigvee A_r(u) \leq u$ we have 
(see \cref{dfn:minsup-max}) 
\[
\sigma(\mu(w,r,s),w,r) \leq s.
\]
However, if $\sigma(\mu(w,r,s),w,r) < s$, then we may find $v = 
\bigvee A_r(v)$ satisfying $\abs*{A_r(v)} \geq \mu(w,r,s)$ and 
$\norm[1]{v} < r+s < (w+1)r$. Therefore, we know that $A_r(v)\neq
\Delta^w_r$, hence there exist $v',v''\in\Delta^w_r$, $v'\not\in
A_r(v)$ (thus $v'\not\leq v$) and $v''\in A_r(v)$, satisfying
$d_1(v',v'')=1$. It follows that $\norm[1]{v\vee v'}=\norm[1]{v}+1
\leq r+s$, in contradiction to $\abs*{A_r(u)}=\mu(w,r,s)$. It follows
that $\sigma(\mu(w,r,s),w,r) = s$.
\end{IEEEproof}

\begin{corollary}\label{cor:mu}
If $\mu(w,r,s)<m\leq \mu(w,r,s+1)$ then
\[
\sigma(m,w,r) = s+1.
\]
\end{corollary}
\begin{IEEEproof}
Firstly, since $m\mapsto\sigma(m,w,r)$ is non-decreasing by 
definition, then by \cref{lem:sigma-mu} 
\begin{align*}
s = \sigma\parenv*{\mu(w,r,s),w,r} \leq \sigma(m,w,r) &\leq \\
\leq \sigma\parenv*{\mu(w,r,s+1),w,r} = s &+ 1.
\end{align*}
However, if $\sigma(m,w,r) = s$, by finding $u_1,\ldots,u_m\in
\Delta^w_r$ with $\norm[1]{\bigvee_{i=1}^m u_i}=r+s$ we deduce 
$\mu(w,r,s)\geq m$, in contradiction.
\end{IEEEproof}

Since we now know that calculating $\mu(w,r,s)$ is sufficient for our
purposes, we turn to that task; since our focus is $\typ^n$, we may do
so for the relevant ranges of $w,r$, whenever that is simpler.

\begin{lemma}\label{lem:spread}
For $w,r,s\in\N$ there exists $u\in\Delta^w_{r+s}$ such that 
$\abs*{A_r(u)} = \mu(w,r,s)$ and for all $1\leq i<j\leq w+1$ it holds 
that $\abs*{u(i)-u(j)}<2$.
\end{lemma}
\begin{IEEEproof}
Take $u\in\Delta^w_{r+s}$ satisfying $\abs*{A_r(u)} = \mu(w,r,s)$, and 
assume to the contrary that there exist $i,j$ such that, w.l.o.g., 
$u(j)\geq u(i)+2$. Denote by $u'$ the vector which agrees on $u$ on 
all coordinates except $u'(j)=u(j)-1$ and $u'(i)=u(i)+1$.

Further, partition $A_r(u)$ and $A_r(u')$ by the projection on the 
subspace formed by all the coordinates except $i$ and $j$. For any 
matching classes $C,C'\subseteq\Delta^w_r$ in the corresponding 
partitions, denote by $t(C)=t(C')$ the difference between $r$ and the 
sum of all coordinates other than $i,j$; Note that $\abs*{C}$ is the 
number of ways to distribute $t(C)$ balls into two bins with 
capacities $u(i),u(j)$ (and correspondingly $u'(i),u'(j)$ for 
$\abs*{C'}$), hence 
\begin{align*}
\abs*{C} &= \min\bracenv*{t(C), u(i)} 
- \max\bracenv*{t(C)-u(j), 0} + 1 \\
&\leq \min\bracenv*{t(C), u(i)+1} 
- \max\bracenv*{t(C)-u(j)+1, 0} + 1 \\
&= \min\bracenv*{t(C'), u'(i)} 
- \max\bracenv*{t(C)-u'(j), 0} + 1 = \abs*{C'},
\end{align*}
where the inequality is justified by cases for $t(C)$, and is strict 
only if $u(i)<t(C)<u(j)$. Thus, the proof is concluded.
\end{IEEEproof}

\cref{lem:spread} allows us to find $\mu(w,r,s)$ with relative 
ease; perhaps the most straightforward example of that is a precise 
calculation for the cases $s=1,2$, which we present next; following 
the examples we conduct a more extensive evaluation, for $s>2$ and the 
relevant ranges of $w,r$.

\begin{example}\label{xmpl:mu_1}
Any vector $u\in\Delta^w_{r+1}$ having $1+\min\bracenv*{w,r}$ positive
coordinates has precisely
\[
\abs*{A_r(u)} = 1+\min\bracenv*{w,r},
\]
since any lower bound in $\Delta^w_r$ is reached by subtracting $1$ 
from a chosen positive coordinate. By \cref{lem:spread} one such 
vector satisfies $\mu(w,r,1) = \abs*{A_r(u)}$, therefore
\[
\mu(w,r,1) = 1+\min\bracenv*{w,r}.
\]
\end{example}

\begin{example}\label{xmpl:mu_2}
We define an injection 
\[
\xi\colon\mathset*{v\in\N^{w+1}}{v\leq u} \to \N^{w+1}
\]
by $\xi(v) \deq u-v$; then clearly, $\xi$ is distance 
preserving, and in particular injective. Hence, 
\[
\mu(w,r,2)\leq \abs*{\Delta^w_2} = \binom{w+2}{2}.
\]
This is achieved with equality when $r+2\geq 2(w+1)$, since there 
exists $(2,2,\ldots,2)\leq v\in \Delta^w_{r+2}$, and it holds that 
$\xi\parenv*{A_r(v)} = \Delta^2_2$). The inequality is strict, 
however, when $r<2w$.

To examine the remaining cases, note first that increasing any 
coordinate of $u$ above $2$ has no effect on $\abs*{A_r(u)}$. Further, 
we again know by \cref{lem:spread} that $\mu(w,r,2)$ is achieved 
when $u$ has the greatest number of positive coordinates, and among 
such vectors, the greatest number greater than or equal to $2$. Now, 
by counting the number of lower bounds for any such $u\in
\Delta^w_{r+2}$ we see that 
\[
\mu(w,r,2) = \begin{cases}
\binom{w+2}{2}, & r\geq 2w; \\
\binom{w+1}{2}+(r-w+1), & w-1\leq r<2w; \\
\binom{r+2}{2}, & r<w-1.
\end{cases}
\]
\end{example}

As can now be seen, a complete evaluation of $\mu(w,r,s)$ for $s>2$ is 
possible using \cref{lem:spread}, but it involves application of 
the inclusion-exclusion principle and its results are not 
illuminating. We shall see instead that an asymptotic evaluation of 
$\mu(w,r,s)$ for typical ranges of $w,r$ will suffice. To do so, we 
note the following proposition.

\begin{lemma}\label{lem:mu_s}
Fix $t$, and take $w,r$ such that $r+t\leq w+1$. For all $s\leq t$ it 
holds that 
\[
\mu(w,r,s) = \binom{r+s}{s}.
\]
\end{lemma}
\begin{IEEEproof}
By \cref{lem:spread} we know that $u\in\Delta^w_{r+s}$ achieving 
$\abs*{A_r(u)} = \mu(w,r,s)$ is such that $r+s$ of its coordinates 
equal $1$, and the remaining $w+1-r-s$ equal $0$. The proposition 
follows.
\end{IEEEproof}

We can use \cref{cor:mu} together with \cref{lem:mu_s} to 
establish the main result of this section, in the following theorem. 
Before doing so, we note a consequence of, e.g., \cref{lem:mu_s}, 
namely that for any string $x\in\typ^n$, and any $y\in D^t(x)$, it 
holds that 
\[
\abs*{\mathset*{x'\in\typ^n}{y\in D^t(x')}} = O(n^t).
\]
Hence, we have for $m_n=\omega(n^t)$ that $N_t(m_n,\typ^n)=o(1)$; it 
is therefore only interesting to find an asymptotic expression for 
$N_t(m_n,\typ^n)$ when $m_n=O(n^t)$.

\begin{theorem}\label{thm:large-m}
Fix $t$ and a sequence $m_n=O(n^t)$. Then 
\begin{align*}
N_t(m_n,\typ^n) 
\sim \tfrac{1}{\parenv*{e_t(m_n,n)}!} 
\parenv*{\tfrac{q-1}{q} n}^{e_t(m_n,n)},
\end{align*}
where $e_t(m_n,n) = t - \ceilenv*{\log_n(m_n)} - \delta(m_n,n)$ and 
$\delta(m,n)\in \bracenv*{0,1}$ is a non-decreasing function in $m$.
\end{theorem}
\begin{IEEEproof}
Let $s \deq \ceilenv*{\log_n(m_n)}$.

Recall from \cref{lem:mu_s} that for $w\geq r+t-1$
\[
\mu(w,r,s-1) = \binom{r+s-1}{r} < \frac{(r+s-1)^{s-1}}{(s-1)!},
\]
hence for $r$ satisfying $\abs*{r-\frac{q-1}{q(q^k-1)}(n-k)} 
< 2 n^{3/4}$ and sufficiently large $n$ 
\[
\log_n \mu(w,r,s-1) < s-1.
\]

On the other hand we have 
\[
\mu(w,r,s+1) = \binom{r+s+1}{r} > \frac{r^{s+1}}{(s+1)!},
\]
and therefore, for such~$r$, 
\begin{align*}
\log_n \mu(w,r,s+1) &> \log_n\parenv*{\frac{1+o(1)}{(s+1)!} 
\parenv*{\frac{q-1}{q(q^k-1)} n}^{s+1}} \\
&= s+1 + o(1).
\end{align*}

Since $s-1 < \log_n(m_n) \leq s$ it now follows from \cref{cor:mu}, 
for sufficiently large~$n$ (which does not depend on~$s$, i.e., 
on~$m_n$), and $w,r$ satisfying 
\begin{align*}
&\abs*{w - \tfrac{q-1}{q}(n-k)} < n^{3/4} \\
&\abs*{r - \tfrac{q-1}{q(q^k-1)}(n-k)} < 2n^{3/4},
\end{align*}
that 
\[
\sigma(m_n,w,r) = s + \delta(m_n,n,r),
\]
where 
\[
\delta(m_n,n,r) = \begin{cases}
1, & m_n > {\textstyle\mu(w,r,s)=\binom{r+\ceilenv*{\log_n(m_n)}}{r}}; \\
0, & \text{otherwise}.
\end{cases}
\]

Next, for such $n,w,r$ we have 
\begin{align*}
&\binom{w+t-\sigma(m_n,w,r)}{w} \\
&\qquad\qquad= \tfrac{1+o(1)}{(t - (s + \delta(m_n,n,r)))!} 
\parenv*{\tfrac{q-1}{q} n}^{t - (s + \delta(m_n,n,r))}.
\end{align*}
It therefore follows from \cref{cor:bar-nobar} and 
\cref{cor:N-by-sup} that 
\begin{align*}
N_t(m_n,\typ^n) &= \tfrac{1+o(1)}{(t - (s + \delta(m_n,n)))!} 
\parenv*{\tfrac{q-1}{q} n}^{t - (s + \delta(m_n,n))} \\
&= \tfrac{1+o(1)}{e_t(m_n,n)!} 
\parenv*{\tfrac{q-1}{q} n}^{e_t(m_n,n)},
\end{align*}
where $\delta(m_n,n)=1$ if and only if $\delta(m_n,n,r)=1$ for all~$r$ 
satisfying the above requirement, and $e_t(m_n,n)$ is as defined in 
the theorem's statement.
\end{IEEEproof}

Finally, we conclude the section by referring back to 
\cref{alg:decode}, proving its validity, and analyzing its 
run-time complexity.
\begin{theorem}\label{thm:decode}
\cref{alg:decode} operates in $O(n^t) = \poly(N)$ steps, and 
produces $x_1,\ldots,x_l\in\typ^n$, $l<m$, such that 
\[
y_1,\ldots,y_{N+1} \in S_t(x_1,\ldots,x_l)\setminus 
\left.\bigcup\right._{\substack{x\in\typ^n \\ 
\mathclap{x\not\in\bracenv*{x_1,\ldots,x_l}}}} D^t(x).
\]
\end{theorem}
\begin{IEEEproof}
First, note that the existence of an ancestor for all
$y_1,\ldots,y_{N+1}$ implies that $y_i\in D^*\parenv*{\rt(y_1)}$ for
all $i$. Moreover, note that finding any $v_i$ may be done in $O(n)$
steps (by calculating $\bar{\phi}(y_i)$ and recording lengths of runs 
of zeros in the process). Any one of these can also produce
$\rt(y_1)$. Hence \hyperref[alg:decode:1]{Step \ref*{alg:decode:1}}
concludes in $O(N n)$ steps.

\hyperref[alg:decode:2]{Step \ref*{alg:decode:2}} can also be
performed in $O(N w) = O(N n)$ steps.

Now, note that since an ancestor of all $y_i$'s exists in $\Sigma^n$,
$r'\geq r$. It is hence possible to compute $A_r(u)$. This may be
achieved by finding all ways of distributing $r'-r < t$ balls into
$w+1$ bins with capacities $u(j)$, e.g., by utilizing combination
generators for all $\binom{w+r'-r}{w}$ combinations, then discarding 
combination which violate the bin-capacity restriction. Combination 
generating algorithms exist which generate all combinations in 
$O\parenv*{\binom{w+r'-r}{w}} = O(n^{t-1})$ steps (e.g., see 
\cite{RusSawWil12}), and pruning illegal combinations can be done in 
$O(w)$ steps each. 
\hyperref[alg:decode:3]{Step~\ref*{alg:decode:3}} can therefore be 
performed in $O(n^t)$ steps.

Finally, the pre-image $\psi_{\rt(y_1)}^{-1}\parenv*{A_r(u)}$ is a set
of ancestors of $y_1, \ldots, y_{N+1}$, which is a subset $\typ^n$,
and no other element of $\typ^n$ is an ancestor of
$y_1,\ldots,y_{N+1}$. We also know that $\abs*{A_r(u)}<m$, otherwise a
contradiction is reached to the definition of $N$. Computing
$\psi_{\rt(y_1)}^{-1}\parenv*{A_r(u)}$ given $\rt(y_1)$ requires
$O(\abs*{A_r(u)} w) \leq O(m n)$ steps.
\end{IEEEproof}

By an examination of the proof, we note that \cref{alg:decode} 
may be applied to list-decode elements of any $C\subseteq\Sigma^n$, 
and not necessarily $\typ^n$; the proof remains unchanged, except that 
one cannot deduce from the existence of a single $x\in C$ such that 
$y_1,\ldots,y_{N+1}\in D^t(x)$, that 
$\psi_{\rt(y_1)}^{-1}\parenv*{A_r(u)}\subseteq C$. Instead, in the 
general case, that verification (and discardment of invalid outputs) 
must be performed as an additional step. This may be unnecessary in 
some cases (e.g., if $C=\Sigma^n$).

\section{Uncertainty with underlying ECC}\label{sec:ecc}

In the previous section, a reconstruction problem with a list-decoding
algorithm was considered, when the underlying message space was
unconstrained (more precisely, constrained only to a typical set). 
However, one is naturally interested in a more general setting, in 
which the message space may be a code with a given minimum distance. 
Thus, in this section, we shall consider the uncertainty associated 
with codes $C\subseteq\typ^n$ such that for all distinct $c,c'\in C$, 
$d(c,c')\geq d$, for some $d>0$. We start with a definition of a 
typical set with a minimum distance.

\begin{definition}
Given $m,n,t,d\in\N$, the \emph{uncertainty associated with the 
minimum distance $d$ in the typical sense} is defined as
\[
N^{\typ}_t(m,n,d) \deq 
\max_{\substack{x_1,\ldots,x_m\in \typ^n \\ d(x_i,x_j)\geq d}} 
\abs*{S_t\parenv*{x_1,\ldots,x_m}}.
\]
Correspondingly, for $w,r\in\N$, 
\begin{align*}
\bar{N}_t(m,w,r,d) &\deq\max_{\substack{u_1,\ldots,u_m\in\Delta^w_r \\ 
d_1(u_i,u_j)\geq d}} \abs*{\bar{S}_t\parenv*{u_1,\ldots,u_m}}, \\
\mu(w,r,s,d) &\deq \max_{u\in\Delta^w_{r+s}} 
\max\mathset*{\abs*{C}}{\begin{smallmatrix}
C\subseteq A_r(u) \\ 
\forall v\neq v'\in C: d_1(v,v')\geq d
\end{smallmatrix}} \\
\sigma(m,w,r,d) 
&\deq \min_{\substack{u_1,\ldots,u_m\in\Delta^w_r \\ 
d_1(u_i,u_j)\geq d}} \norm[1]{\left.\bigvee\right._{i=1}^m u_i} - r.
\end{align*}
\end{definition}

It should be noted that if $d>t$ then $N^{\typ}(2,n,d)=0$, meaning 
that unique decoding from a single noisy output is possible. It was 
seen in \cite{YehSch19} that $d=t$ suffices for unique reconstruction 
($m=2$) with sub-linear uncertainty (in fact, $N=1$, which corresponds 
to receiving two distinct noisy outputs, suffices). We shall 
incidentally see that again while considering $d\leq t$.

As in the previous section, we defer study of the uncertainty 
$N^{\typ}_t(m,n,d)$, and begin by presenting a list-decoding scheme 
given sufficiently many ($N^{\typ}_t(m,n,d)+1$) distinct strings in 
\[
D^t(C) \deq \bigcup_{c\in C} D^t(c),
\]
for some given code $C\subseteq\typ^n$ with minimum distance $d$. We 
shall assume that a decoding scheme for recovering from at most $d-1$ 
errors is known for $C$, which we denote by 
$\cD\colon\Sigma^{n+k(d-1)} \to C$.

\begin{myalgorithm}\label{alg:decode-ecc}
Fix $n,m$ and $d\leq t$; take $C\subseteq \typ^n$ with 
minimum $d(\cdot,\cdot)$ distance $d$ (see \cref{def:dist}), and 
assume a decoding scheme for recovering up to $d-1$ 
tandem-duplication errors is provided. 
Denote $N\deq N^{\typ}_t(m,n,d)$ and assume as input distinct 
$y_1,\ldots,y_{N+1} \in \Sigma^{n+kt}$ such that there exists $x\in C$ 
satisfying $y_1,\ldots,y_{N+1} \in D^t(x)$.
\begin{enumerate}
\item\label{alg:decode-ecc:1}
Apply \cref{alg:decode} to obtain $z_1,\ldots,z_l\in 
\Sigma^{n+k(d-1)}$ such that 
\[
y_1,\ldots,y_{N+1} \in S_{t-d+1}(z_1,\ldots,z_l)\setminus 
\bigcup_{\mathclap{\qquad\qquad\substack{z\in\Sigma^{n+k(d-1)} \\ 
z\not\in\bracenv*{z_1,\ldots,z_l}}}} D^{t-d+1}(z).
\]

\item\label{alg:decode-ecc:2}
Decode each $z_i$ with the provided algorithm to produce $x_i \deq 
\cD(z_i)\in C$; if $z_i\not\in D^{d-1}(x_i)$, discard $x_i$.

\item\label{alg:decode-ecc:3}
Return every $x_i$ that was not discarded in the last step, as a list.
\end{enumerate}
\end{myalgorithm}

Proof of correctness and analysis of run-time complexity will be 
presented in \cref{thm:decode-ecc}, at the end of the section. At 
this point we will instead present an example of the algorithm's 
application.
\begin{example}
We continue the discussion of \cref{xmpl:alg:decode}. In 
particular, we let $q=3, k=2$, but this time take $n=9, t=4$. As 
before, we shall read multiple distinct elements of $D^t(x)$, where 
$x$ is now an unknown element of a code $C\subseteq \typ^n$ correcting 
a single tandem-duplication (i.e., with minimum $d(\cdot,\cdot)$ 
distance $d=2$). We would like to decode a list of strings in $C$, of 
which $x$ is a member. We also make the arbitrary decision to require 
a list of size at most $2$, setting $m=3$ (which may be justified by 
our desire to do at least as well as we did in the previous example, 
since the added redundancy of an error-correcting code can be expected 
to offset the additional duplication error we allowed for; the more 
cynical reader will note that it is also easier to analyze).

We make the same first read as in \cref{xmpl:alg:decode}:
\[
y_1 = 10101012122222222.
\]
(Recall, $\rt(y_1)=\rt(x)=10122$.)

We still have $w=\wt_H(\bar{\phi}(y_1))=2$, but in this case 
$r = \norm[1]{\psi_{\rt(y_1)}\parenv*{\bar{\phi}(y_1)}} - t = 2$. 
Now, it suffices to obtain $2$ distinct reads.

(Again, a reader looking back at this example will note that we may 
use Lem.~23 to determine--after a short exhaustive search--that 
$\mu(w,r,3,d)=2$ and $\mu(w,r,4,d)=3$. This implies, by Lem.~22, that 
$\sigma(m,w,r,d)=4$, and \cref{cor:N-by-sup-ecc} determines that 
$\bar{N}_t(m,w,r,d) = 1$.)

We therefore make one other distinct read; we'll use $y_2$ from 
\cref{xmpl:alg:decode},
\[
y_2 = 10101010122222222,
\]
and we've already seen that 
\begin{align*}
v_1 &= \psi_{\rt(y_1)}(y_1) = (2,1,3),\\
v_2 &= \psi_{\rt(y_1)}(y_2) = (3,0,3).
\end{align*}
To conclude the application of \cref{alg:decode} in 
\hyperref[alg:decode-ecc:1]{Step \ref*{alg:decode-ecc:1}}, we find 
$u = v_1\wedge v_2 = (2,0,3)$, and $A_{r+d-1}(u)=A_3(u) = 
\bracenv*{u_1, u_2, u_3}$, where
\begin{align*}
u_1 &= (1,0,2),\\
u_2 &= (2,0,1),\\
u_3 &= (0,0,3).
\end{align*}

Denoting $z_i=\psi_{\rt(y_1)}^{-1}(u_i)$, $i\in\bracenv*{1,2,3}$, we 
might now find 
\begin{align*}
\phi(z_1) &= 10000210000,\\
\phi(z_2) &= 10000002100,\\
\phi(z_3) &= 10021000000,
\end{align*}
and consequently
\begin{align*}
z_1 &= 10101222222,\\
z_2 &= 10101012222,\\
z_3 &= 10122222222.
\end{align*}
However, 
\hyperref[alg:decode-ecc:2]{Step \ref*{alg:decode-ecc:2}} calls for 
an application of the decoding function $\cD$, which is more easily 
conceptualized in $\N^3$. 
Recall, $\psi_{\rt(y_1)}\parenv*{C\cap D^*(\rt(y_1)}\subseteq 
\Delta^w_r = \Delta^2_2$, and has minimum $d_1$ distance $2$ 
(correcting a single tandem duplication). It may be the following 
(optimal) such code:
\[
\bracenv*{(2,0,0),(0,2,0),(0,0,2)},
\]
in which case the decoder outputs
\begin{align*}
u'_1 &\deq \cD(u_1) = (0,0,2),\\
u'_2 &\deq \cD(u_1) = (2,0,0),\\
u'_3 &\deq \cD(u_1) = (0,0,2) = u'_1.
\end{align*}
Having $x'_i=\psi_{\rt(y_1)}^{-1}(u'_i)$, $i\in \bracenv*{1,2}$, we 
find
\begin{align*}
\phi(x'_1) &= 100210000,\\
\phi(x'_2) &= 100000021,
\end{align*}
and therefore
\begin{align*}
x'_1 &= 101222222,\\
x'_2 &= 101010122.
\end{align*}
Clearly, $z_i\in D^{d-1}(x_i) = D^1(x_i)$, $i\in \bracenv*{1,2}$, 
hence \cref{alg:decode-ecc} is concluded.

We might remark, however, that it is also possible that the underlying 
error-correcting code being used satisfies, e.g., 
\[
\psi_{\rt(y_1)}\parenv*{C\cap D^*(\rt(y_1)} = \bracenv*{(2,0,0),
(0,1,1)},
\]
in which case $u_1,u_3\not\in D^1\parenv*{C\cap D^*(\rt(y_1)}$; hence,
we cannot know $u''_1 = \cD(u_1)$ and $u''_3 = \cD(u_3)$. It is 
possible that either equals $u''_2 = \cD(u_2) = (2,0,0)$, but it is 
just as feasible (depending on the chosen implementation of $\cD$) 
that either might be $(0,1,1)$.

Nevertheless, we may verify (concluding 
\hyperref[alg:decode-ecc:2]{Step \ref*{alg:decode-ecc:2}}) that 
$z_2\in D^1(x''_2)$ (where we denote $x''_i = 
\psi_{\rt(y_1)}^{-1}(u''_i)$, $i\in \bracenv*{1,2,3}$), but 
$z_1,z_3\not\in D^1((0,1,1))$, hence after discarding invalid outputs 
we output only $x''_2$.
\end{example}

Next, we show that $N^{\typ}_t(m,n,d)$ may be analyzed in terms of 
$\bar{N}_t(m,w,r,d)$.
\begin{corollary}\label{cor:bar-nobar-ecc}
For all sufficiently large $n$,
\begin{align*}
& N^{\typ}_t(m,n,d) =\\
&\quad = \max\mathset*{\bar{N}_t(m,w,r,d)}{
\begin{smallmatrix}
\abs*{w-\frac{q-1}{q}(n-k)} < n^{3/4} \\
\abs*{r-\frac{q-1}{q(q^k-1)}(n-k)} < 2 n^{3/4}
\end{smallmatrix}
}.
\end{align*}
\end{corollary}
\begin{IEEEproof}
Similarly to the proof of \cref{lem:intrsct-smplx}, a choice of 
$x_1,\ldots,x_m\in \typ^n$ satisfying $d(x_i,x_j)\geq d$ and 
$\abs*{S_t(x_1,\ldots,x_m)}=N^{\typ}_t(m,n,d)$ must also satisfy 
$x_i\sim_k x_j$ (otherwise $S_t(x_1,\ldots,x_m)=\emptyset$), hence we 
may find $x\deq \rt(x_1) = \ldots = \rt(x_m)$. In addition 
\[
\abs*{\bar{S}_t\parenv*{\psi_x(x_1),\ldots,\psi_x(x_m)}} 
= \abs*{S_t(x_1,\ldots,x_m)}
\]
and $d_1\parenv*{\psi_x(x_i),\psi_x(x_j)}\geq d$.
The other direction follows as in the proof of 
\cref{cor:bar-nobar}.
\end{IEEEproof}

We shall continue using an analogous approach to that of the previous 
section, in finding $\bar{N}_t(m,w,r,d)$ in order to estimate 
$N^{\typ}_t(m,n,d)$.

\begin{corollary}\label{cor:N-by-sup-ecc}
$\bar{N}_t(m,w,r,d) = \binom{w+t-\sigma(m,w,r,d)}{w}$.
\end{corollary}
\begin{IEEEproof}
This proposition follows from \cref{lem:intersection-sup} in 
similar fashion to \cref{cor:N-by-sup}.
\end{IEEEproof}

\begin{lemma}\label{lem:sigma-mu-ecc}
Take some $m,w,r,s,d\in\N$. If 
\[
\mu(w,r,s,d)<m\leq \mu(w,r,s+1,d)
\]
then
\[
\sigma(m,w,r,d) = s+1.
\]
\end{lemma}
\begin{IEEEproof}
The proof follows the same arguments as in the proofs of 
\cref{lem:sigma-mu} and \cref{cor:mu}.
\end{IEEEproof}

\begin{lemma}\label{lem:spread-ecc}
For $0<w,r,s,d\in\N$ there exist $u\in\Delta^w_{r+s}$, and 
$C\subseteq A_r(u)$ with minimum $d_1$ distance $d$, satisfying 
$\abs*{C} = \mu(w,r,s,d)$, such that for no pair $1\leq i,j\leq w+1$, 
$i\neq j$, it holds that $u(i)\geq 2$ and $u(j)=0$.
\end{lemma}
\begin{IEEEproof}
Take $u\in\Delta^w_{r+s}$ and $C\subseteq A_r(u)$ satisfying 
$\abs*{C} = \mu(w,r,s,d)$, and assume to the contrary that there exist 
such $i,j$; denote by $u'$ the vector which agrees with $u$ on all 
coordinates except $u'(j)=1$ and $u'(i)=u(i)-1$. The proposition is 
justified by finding any isometric injection~$\rho\colon A_r(u) 
\to A_r(u')$.

Indeed, define $\rho(v) \deq v$ if $v(i)<u(i)$, otherwise 
\[
\parenv*{\rho(v)}(l) \deq \begin{cases}
u(i)-1, & l=i; \\
1, & l=j; \\
v(l), & \text{otherwise}.
\end{cases}
\]
Then $\rho$ is well defined. 
Moreover, take any $v_1,v_2\in A_r(u)$. If $v_1(i),v_2(i)<u(i)$ then 
clearly $d_1(\rho(v_1),\rho(v_2)) = d_1(v_1,v_2)$. The same trivially 
holds when $v_1(i) = v_2(i) = u(i)$. If, w.l.o.g. $v_1(i) < v_2(i) = 
u(i)$, then 
\begin{align*}
\abs*{\parenv*{\rho(v_1)}(i)-\parenv*{\rho(v_2)}(i)} 
&= \abs*{v_1(i)-\parenv*{\rho(v_2)}(i)} \\
&= \abs*{v_1(i)-v_2(i)} - 1
\end{align*}
but 
\begin{align*}
\abs*{\parenv*{\rho(v_1)}(j)-\parenv*{\rho(v_2)}(j)} 
&= \abs*{v_1(j)-\parenv*{\rho(v_2)}(j)} = \abs*{0 - 1} \\
&= 1 = \abs*{v_1(j)-v_2(j)} + 1,
\end{align*}
hence, once again, $d_1(\rho(v_1),\rho(v_2)) = d_1(v_1,v_2)$.
\end{IEEEproof}

As in \cref{sec:typ}, \cref{lem:spread-ecc} allows us to find 
$\mu(w_n,r_n,s)$ for typical ranges of $w_n,r_n$, using binary 
constant-weight codes. 
This is given precise meaning in the following definition and lemma.

\begin{definition}
Denote by~$\GF(2)$ the field of size 2, and by~$d_H$ the Hamming 
metric. Denote by $A(\nu,2\delta,\omega)$ the size of the largest 
length~$\nu$ binary code with minimum Hamming distance~$2\delta$ and 
constant Hamming weight~$\omega$.
\end{definition}

\begin{lemma}\label{lem:mu_s-ecc}
Fix $t$, and take $w,r$ such that $r+t\leq w+1$. For all $s\leq t$ it 
holds that 
\[
\mu(w,r,s,d) = A\parenv*{r+s,2d,s}.
\]
\end{lemma}
\begin{IEEEproof}
By \cref{lem:spread-ecc} we know that there exist 
$u\in\Delta^w_{r+s}$ and $C\subseteq A_r(u)$ satisfying 
\begin{itemize}
\item
$\abs*{C}=\mu(w,r,s,d)$.

\item
For all $v_1,v_2 \in C$, $v_1\neq v_2$, it holds that $d_1(v_1,v_2)
\geq d$.

\item
$u$ has $r+s$ of its coordinates equal $1$, and the remaining 
$w+1-r-s$ equal $0$.
\end{itemize}

Define $\rho\colon A_r(u)\to \GF(2)^{r+s}$ by restricting $u-v$ to the 
support of $u$ (and identifying $\GF(2)$ with $\bracenv*{0,1}\subseteq
\N$). Then $\rho$ is a bijection onto constant-Hamming-weight~$s$ 
elements of $\GF(2)^{r+s}$. Further, for all $v_1,v_2\in A_r(u)$ it 
holds that 
\[
d_H(\rho(v_1),\rho(v_2)) = 2 d_1(v_1,v_2).
\]
Hence, there's a size-preserving one-to-one correspondence between 
codes $C'\subseteq A_r(u)$ with minimum $d_1$ distance $d$, and codes 
in $\GF(2)^{r+s}$ with minimum Hamming distance~$2d$ and constant 
Hamming weight~$s$. The proposition follows.
\end{IEEEproof}

We can now summarize our observations in the following theorem.
\begin{theorem}\label{thm:large-m-ecc}
Fix $d\leq t$ and a sequence $m_n=O(n^{t-d+1})$. Then 
\begin{align*}
N^{\typ}_t(m_n,n,d) 
\sim \tfrac{1}{\parenv*{e_t(m_n,n,d)}!} 
\parenv*{\tfrac{q-1}{q} n}^{e_t(m_n,n,d)},
\end{align*}
where $e_t(m_n,n,d) = t - \ceilenv*{\log_n(m_n)} - d + 
\epsilon(m_n,n,d)$ and $\epsilon(m,n,d)\in \bracenv*{0,1}$ is a 
non-increasing function of $m$.
\end{theorem}
\begin{IEEEproof}
The proof follows the same lines as that of \cref{thm:large-m}. 
Let $s \deq \ceilenv*{\log_n(m_n)}+d-1$.

Recall from the first Johnson bound \cite[Th.~2]{Joh62} that 
\begin{align*}
A(r+s-1,2d,s-1) &\leq \binom{r+s-1}{s-d} \bigg/ \binom{s-1}{s-d} \\
&< \frac{(d-1)!}{(s-1)!} (r+s-1)^{s-d},
\end{align*}
hence for $r$ satisfying $\abs*{r-\frac{q-1}{q(q^k-1)}(n-k)} 
< 2 n^{3/4}$ and sufficiently large $n$ 
\[
\log_n A(r+s-1,2d,s-1) < s-d.
\]

On the other hand, by \cite[Th.~6]{GraSlo80} we have 
\[
A(r+s+1,2d,s+1)\geq \frac{1}{p^{d-1}} \binom{r+s+1}{s+1}
\]
for any prime power~$p$, $p > r+s$. By the prime number theorem (a 
weaker version, or even Bertrand's postulate, suffices. See, e.g., 
\cite{Che1852}) there exists in fact such prime number~$p$ satisfying 
$r+s < p \leq n$ for sufficiently large~$n$ and $r$ satisfying 
$\abs*{r-\frac{q-1}{q(q^k-1)}(n-k)} < 2 n^{3/4}$, hence in particular 
\begin{align*}
A(r+s+1,2d,s+1) &\geq \frac{1}{n^{d-1}} \binom{r+s+1}{s+1} \\
&> \frac{r^{s+1}}{n^{d-1} (s+1)!},
\end{align*}
and therefore
\begin{align*}
&\log_n A(r+s+1,2d,s+1) \\
&\quad> \log_n\parenv*{\frac{1+o(1)}{n^{d-1} (s+1)!} 
\parenv*{\frac{q-1}{q(q^k-1)} n}^{s+1}} \\
&\quad= s-d+2 + o(1).
\end{align*}

Since $s-d < \log_n(m_n) \leq s-d+1$ it now follows from 
\cref{lem:sigma-mu-ecc} and \cref{lem:mu_s-ecc}, for 
sufficiently large~$n$ (which does not depend on~$s$, i.e., on~$m_n$), 
and $w,r$ satisfying 
\begin{align*}
&\abs*{w - \tfrac{q-1}{q}(n-k)} < n^{3/4} \\
&\abs*{r - \tfrac{q-1}{q(q^k-1)}(n-k)} < 2n^{3/4},
\end{align*}
that 
\[
\sigma(m_n,w,r,d) = s + \delta(m_n,n,r,d),
\]
where 
\[
\delta(m_n,n,r,d) = \begin{cases}
1, & m_n > A(r+s,2d,s); \\
0, & \text{otherwise}.
\end{cases}
\]
(Note that that $s$ is a function of~$m_n,n$.)

Next, for such $n,w,r$ we have 
\begin{align*}
&\binom{w+t-\sigma(m_n,w,r,d)}{w} \\
&\qquad\qquad= \tfrac{1+o(1)}{(t - (s + \delta(m_n,n,r,d)))!} 
\parenv*{\tfrac{q-1}{q} n}^{t - (s + \delta(m_n,n,r,d))}.
\end{align*}
It therefore follows from \cref{cor:bar-nobar-ecc} and 
\cref{cor:N-by-sup-ecc} that 
\begin{align*}
N^{\typ}_t(m_n,n,d) &= \tfrac{1+o(1)}{(t - (s + \delta(m_n,n,d)))!} 
\parenv*{\tfrac{q-1}{q} n}^{t - (s + \delta(m_n,n,d))} \\
&= \tfrac{1+o(1)}{e_t(m_n,n,d)!} 
\parenv*{\tfrac{q-1}{q} n}^{e_t(m_n,n,d)},
\end{align*}
where $\delta(m_n,n,d)=1$ if and only if $\delta(m_n,n,r,d)=1$ for 
all~$r$ satisfying the above requirement, $\epsilon(m_n,n,d) 
\deq 1-\delta(m_n,n,d)$, and $e_t(m_n,n,d)$ is as defined in the 
theorem's statement.
\end{IEEEproof}

It is again remarked here that in the case that coding is performed 
with $d=t$, we observe that unique reconstruction ($m=2$) is possible 
with just two reads ($N=1$); To see that, note that $\delta(2,r,d)=0$ 
for all $r\geq d$, hence for sufficiently large $n$ we have 
$\epsilon(2,d)=1$ and therefore $e_t(2,n,d)=0$. This result, as 
mentioned above, was already observed in \cite{YehSch19}.

The trade-off established in \cref{thm:large-m-ecc} between the 
code minimum distance~$d$ (equivalently, its redundancy, since as seen 
in \cite{KovTan18a,LenJunWac18b} and mentioned above, a code with 
minimum distance~$d$ has optimal redundancy~$(d-1) \log_q(n) + O(1)$), 
the number of tandem-duplication errors~$t$, the decoded list 
size~$m_n$, and the resulting uncertainty~$N^{\typ}_t(m_n,n,d)$, is 
perhaps better visualized in the following corollary.

\begin{corollary}\label{cor:tradeoff}
Fix $d\leq t$ and a sequence $m_n=O(n^{t-d+1})$. Then 
\begin{align*}
\log_n N^{\typ}_t(m_n,n,d) + \ceilenv*{\log_n(m_n)} + d &= \\
=&\; t + \epsilon(m_n,d) + o(1),
\end{align*}
where $\epsilon(m,d)\in \bracenv*{0,1}$ is a non-increasing function 
of $m$.
\end{corollary}

Finally, we conclude the section by proving correctness for \cref{alg:decode-ecc}, and analyzing its run-time complexity.
\begin{theorem}\label{thm:decode-ecc}
\cref{alg:decode-ecc} produces $x_1,\ldots,x_{l'}\in C$, $l'<m$, 
such that 
\[
y_1,\ldots,y_{N+1} \in S_t(x_1,\ldots,x_{l'})\setminus 
\left.\bigcup\right._{\substack{x\in\typ^n \\ 
\mathclap{x\not\in\bracenv*{x_1,\ldots,x_l}}}} D^t(x).
\]
Further, it operates in $O(n^t+ n^{t-d+1}\cC)$ steps, where $\cC$ is 
the run-time complexity of $\cD$.
\end{theorem}
\begin{IEEEproof}
There is one assumption to \cref{alg:decode} and 
\cref{thm:decode} which may now not be satisfied, that indeed there 
exists $z\in\Sigma^{n+k(d-1)}$ such that $y_1,\ldots,y_{N+1} \in 
D^t(z)$. If there does not, then 
\hyperref[alg:decode-ecc:1]{Step \ref*{alg:decode-ecc:1}} might fail 
because \cref{alg:decode} finds $\hat{z}\deq \bigwedge_{i=1}^N y_i$ 
with $\abs*{\hat{z}} = n+ks$ and $s<(d-1)$. If that is the case, 
however, such $\hat{z}$ may still be passed on to the next step, since 
we may still decode it to a unique $x\in C$ for which $z\in D^s(x)$ 
(since $C$ has minimum distance $d$, there cannot exist two distinct 
ancestors of $z$ in $C$), which justifies the claim. Otherwise, 
\cref{thm:decode} proves that the first step produces what is 
claimed, and we may assume w.l.o.g. that 
$
s \geq d-1
$.

This assumption now implies that for each $x\in C$ such that 
$y_1,\ldots,y_{N+1} \in D^t(x)$ there exists $z\in D^{d-1}(x)$ such 
that $y_1,\ldots,y_{N+1} \in D^t(z)$, hence $z\in 
\bracenv*{z_1,\ldots,z_l}$; this is because one may arbitrarily choose 
such $x\leq z\leq\hat{z}$. On the other hand, each $z\in 
\Sigma^{n+k(d-1)}$ can be decoded to at most a single $x\in C$ for 
which $z\in D^{d-1}(x)$ (again, due to the code's minimum distance), 
and that $x$ satisfies $y_1,\ldots,y_{N+1} \in D^t(x)$. 
We remark that it is possible that the first step produces $z_i\not\in 
D^{d-1}(C)$, hence $x_i = \cD(z_i)$ may be erroneous (as the decoder 
receives invalid input); however, as $y_1,\ldots,y_{N+1} \in 
D^{t-d+1}(z_i)$, such results can indeed be discarded by testing if 
$z_i \in D^{d-1}(x_i)$.

Note that if distinct $x_1,\ldots,x_m \in C$ are produced by 
\hyperref[alg:decode-ecc:2]{Step \ref*{alg:decode-ecc:2}}, we have 
$
\abs*{S_t(x_1,\ldots,x_m)}\geq\abs*{\bracenv*{y_1,\ldots,y_{N+1}}} 
= N+1
$
and therefore a contradiction. Hence, $l'<m$.

Finally, we know that 
\hyperref[alg:decode-ecc:1]{Step \ref*{alg:decode-ecc:1}} operates 
in $O(n^t) = \poly(N)$ steps. 
Since testing whether $z_i\in D^{d-1}(x_i)$ may be done in $O(n)$ 
steps, \hyperref[alg:decode-ecc:2]{Step \ref*{alg:decode-ecc:2}} 
clearly operates in $O(l(\cC+n))$ steps. Hence, it now suffices to 
show that $l=O(n^{t-d+1})$ to conclude the proof.

To that end, note that the number of
$t$-ancestors of $y\in \Sigma^{n+kt}$ is bound from above by
$\mu(w,r,t)$, where $w = \wt_H(\bar{\phi}(y)) \leq n-k$ and $r =
\norm[1]{\psi_{\rt(y)}(y)} - t$. As in \cref{xmpl:mu_2}, using
$\xi$ we note that
\begin{align*}
\mu(w,r,t) &\leq \abs*{\Delta^w_t} = \binom{w+t}{w}
< \frac{1}{t!}(w+t)^t\\
&\leq \frac{1}{t!}(n-k+t)^t < (n+t)^t.
\end{align*}
Hence
$N_t\parenv*{(n+t)^t ,\typ^n} = 0$; this in particular implies that
for $\hat{m} \deq \parenv*{n+t+(k-1)(d-1)}^{t-d+1}$ we have
\[
N_{t-d+1}\parenv*{\hat{m} ,\typ^{n+k(d-1)}} = 0 
\leq N^{\typ}_t(m,n,d).
\]
Note, then, that $l<\hat{m}$. This result can be considerably improved 
by noting that for all $m'$ satisfying 
\[
N_{t-d+1}(m',\typ^{n+k(d-1)})\leq N^{\typ}_t(m,n,d)
\]
it holds that $l<m'$, but for our purposes $\hat{m}$ does suffice.
\end{IEEEproof}

\appendix[Conclusion of proof of 
\texorpdfstring{\cref{lem:typ}}{Lemma 2}]\label{app:r}

As in the proof of \cref{lem:typ}, we define $u(i)\deq
\parenv*{\bar{\phi}(x)}(i)$.
Further define for all $1\leq i\leq n-k$ and $1\leq j<n-k-i+1$ the 
indicator $I_i(j)$ of the event of a run of precisely $j$ zeros 
starting in $u$ at index $i$. Then 
\begin{align*}
\E[r(x)] =& \sum_{i=1}^{n-k} \sum_{j=1}^{n-k-i+1} 
\floorenv*{\tfrac{j}{k}} \Pr\parenv*{I_i(j)=1} \\
=&\, \floorenv*{\tfrac{n-k}{k}}\Pr\parenv*{I_1(n-k)=1} 
+ \sum_{j=1}^{\mathclap{n-k-1}} \floorenv*{\tfrac{j}{k}} 
\Pr\parenv*{I_1(j)=1} \\
&+ \sum_{i=2}^{n-k} \floorenv*{\frac{n-k-i+1}{k}} 
\Pr\parenv*{I_i(n-k-i+1)=1} \\
&+ \sum_{i=2}^{n-k-1}\sum_{j=1}^{n-k-i} \floorenv*{\tfrac{j}{k}} 
\Pr\parenv*{I_i(j)=1} \\
=&\, \floorenv*{\frac{n-k}{k}} \frac{1}{q^{n-k}} 
+\sum_{j=1}^{n-k-1} \floorenv*{\frac{j}{k}} \frac{q-1}{q^{j+1}} \\
&+ \sum_{i=2}^{n-k} \floorenv*{\frac{n-k-i+1}{k}} 
\frac{q-1}{q^{n-k-i+2}} \\
&+ \sum_{i=2}^{n-k-1} \sum_{j=1}^{n-k-i} \floorenv*{\frac{j}{k}} 
\frac{(q-1)^2}{q^{j+2}} \\
=&\, \frac{\floorenv*{n/k}-1}{q^{n-k}} 
+ 2\frac{q-1}{q}\sum_{j=1}^{n-k-1} \frac{\floorenv*{j/k}}{q^j} \\
&+ \frac{(q-1)^2}{q^2}\sum_{i=2}^{n-k-1} \sum_{j=1}^{n-k-i} 
\frac{\floorenv*{j/k}}{q^j} 
\end{align*}

We note that 
\begin{align*}
\sum_{j=1}^{p} \frac{\floorenv*{j/k}}{q^j} 
=&\, \sum_{j=k}^{p} \frac{\floorenv*{j/k}}{q^j} \\
=&\, \sum_{j=k\floorenv*{p/k}}^p \frac{\floorenv*{p/k}}{q^j} 
+ \sum_{i=1}^{\floorenv*{p/k}-1} \sum_{j=0}^{k-1}\frac{i}{q^{ik+j}} \\
=&\, \frac{q}{q-1} \bigg[\floorenv*{p/k} 
\parenv*{\frac{1}{q^{k\floorenv*{p/k}}} - \frac{1}{q^{p+1}}} \\
&+ \parenv*{1-\frac{1}{q^k}} \sum_{i=1}^{\floorenv*{p/k}-1} 
\frac{i}{q^{ik}} \bigg] \\
=&\, \frac{q}{q-1} \bigg[\floorenv*{p/k} 
\parenv*{\frac{1}{q^{k\floorenv*{p/k}}} - \frac{1}{q^{p+1}}} \\
&+ \frac{1}{q^k-1} \parenv*{1 - \frac{1}{q^{k(\floorenv*{p/k}-1)}}} \\
&- \frac{\floorenv*{p/k}-1}{q^{k \floorenv*{p/k}}} 
\bigg] \\
=&\, \frac{q}{q-1} \bigg[
\frac{1}{q^k-1} \parenv*{1 - \frac{1}{q^{k(\floorenv*{p/k}-1)}}} \\
&+ \frac{1}{q^{k \floorenv*{p/k}}} - \frac{\floorenv*{p/k}}{q^{p+1}}
\bigg] 
\end{align*}

Now 
\[
\frac{\floorenv*{n/k}-1}{q^{n-k}} 
+ 2\frac{q-1}{q}\sum_{j=1}^{n-k-1} \frac{\floorenv*{j/k}}{q^j} = O(1).
\]
Hence, it suffices to find 
\begin{align*}
\frac{(q-1)^2}{q^2} &\sum_{i=2}^{n-k-1} \sum_{j=1}^{n-k-i} 
\frac{\floorenv*{j/k}}{q^j} 
= \frac{(q-1)^2}{q^2} \sum_{p=1}^{n-k-2} \sum_{j=1}^p 
\frac{\floorenv*{j/k}}{q^j} \\
&= \frac{q-1}{q(q^k-1)} \sum_{p=1}^{n-k-2} 
\parenv*{1 - \frac{1}{q^{k(\floorenv*{p/k}-1)}}} \\
&\quad - \frac{q-1}{q} \sum_{p=1}^{n-k-2} \bigg[
\frac{\floorenv*{p/k}}{q^{p+1}} - \frac{1}{q^{k \floorenv*{p/k}}}
\bigg].
\end{align*}
Again, note that $\sum_{p=1}^{n-k-2} \frac{\floorenv*{p/k}}{q^{p+1}} 
= O(1)$; in addition, we note that $\sum_{p=1}^{n-k-2} 
\frac{1}{q^{k \floorenv*{p/k}}} = O(1)$.

We therefore find $\E[r(x)] = \frac{q-1}{q(q^k-1)}(n-k) + O(1)$.

\section*{Acknowledgments}

The authors gratefully acknowledge the two anonymous reviewers and 
associate editor, whose careful reading and suggestions helped shape 
this paper. 
We also offer special thanks to Prof. Jehoshua Bruck for his 
illuminating insight, and the conversation that originally turned our 
attention to the problem explored in this paper.
\begin{IEEEbiographynophoto}{Yonatan Yehezkeally}
(S'12--M'20)
is a postdoctoral researcher in the Associate Professorship of Coding 
and Cryptography (Prof. Wachter-Zeh), Institute for Communications 
Engineering, TU Munich Department of Electrical and Computer 
Engineering. 
His research interests include coding for DNA storage, combinatorial 
structures, algebraic coding, and finite group theory.

Yonatan received the B.Sc.~(\emph{cum laude}) degree in Mathematics, 
and the M.Sc.~(\emph{summa cum laude}) and Ph.D. degrees in Electrical 
and Computer Engineering, in 2013, 2017 and 2020 respectively, all 
from Ben-Gurion University of the Negev, Beer-Sheva, Israel.
\end{IEEEbiographynophoto}

\begin{IEEEbiographynophoto}{Moshe Schwartz}
(M'03--SM'10)
is a professor at the School of Electrical and Computer Engineering, 
Ben-Gurion University of the Negev, Israel. His research interests 
include algebraic coding, combinatorial structures, and digital 
sequences.

Prof.~Schwartz received the B.A.~(\emph{summa cum laude}), M.Sc., and 
Ph.D.~degrees from the Technion -- Israel Institute of Technology, 
Haifa, Israel, in 1997, 1998, and 2004 respectively, all from the 
Computer Science Department. He was a Fulbright post-doctoral 
researcher in the Department of Electrical and Computer Engineering, 
University of California San Diego, and a post-doctoral researcher in 
the Department of Electrical Engineering, California Institute of 
Technology. While on sabbatical 2012--2014, he was a visiting 
scientist at the Massachusetts Institute of Technology (MIT).

Prof.~Schwartz received the 2009 IEEE Communications Society Best 
Paper Award in Signal Processing and Coding for Data Storage, and the 
2020 NVMW Persistent Impact Prize. He has also been serving as an 
Associate Editor for Coding Techniques for the IEEE Transactions on 
Information Theory since 2014, and an Editorial Board Member for the 
Journal of Combinatorial Theory Series A since 2021.
\end{IEEEbiographynophoto}
\end{document}